\documentclass[10pt,conference]{IEEEtran}
\IEEEoverridecommandlockouts

\usepackage{cite}
\usepackage{amsmath,amssymb}
\usepackage{graphicx}
\usepackage{xcolor}
\usepackage{booktabs}
\usepackage{enumitem}
\usepackage{hyperref}
\usepackage{array}
\usepackage{bbm}
\usepackage{comment}
\usepackage{float}

\usepackage{tikz}
\usetikzlibrary{arrows.meta,positioning,shapes.geometric}

\newcommand{\EdgeAI}{Edge AI}
\newcommand{\AdaptiveEdgeAI}{Adaptive Edge AI}
\newcommand{\ASE}{agent-system-environment}

\newcommand{\boxedpara}[1]{%
\vspace{0.3em}
\noindent\fcolorbox{black!25}{black!3}{%
\begin{minipage}{0.97\linewidth}
\vspace{0.35em}
#1
\vspace{0.35em}
\end{minipage}}%
\vspace{0.3em}
}

\begin{document}

\title{Position Paper: From Edge AI to Adaptive Edge AI}

\author{
\IEEEauthorblockN{Fabrizio Pittorino and Manuel Roveri}
\IEEEauthorblockA{Department of Electronics, Information and Bioengineering (DEIB), Politecnico di Milano, Italy\\
Email: \{fabrizio.pittorino, manuel.roveri\}@polimi.it}
}

\maketitle

\begin{abstract}
Edge AI is often framed as model compression and deployment under tight constraints. We argue a stronger operational thesis: \emph{Edge AI in realistic deployments is necessarily adaptive}. In long-horizon operation, a fixed (non-adaptive) configuration faces a fundamental failure mode: as data and operating conditions  evolve and change in time, it must either (i)~violate time-varying budgets (latency/energy/thermal/connectivity/privacy) or (ii)~lose predictive reliability (accuracy and, critically, calibration), with risk concentrating in transient regimes and rare time intervals rather than in average performance. If a deployed system cannot \emph{reconfigure} its computation - and, when required, its model state - under evolving conditions and constraints, it reduces to static embedded inference and cannot provide sustained utility. This position paper introduces a minimal Agent-System-Environment (ASE) lens that makes adaptivity precise at the edge by specifying (i)~what changes, (ii)~what is observed, (iii)~what can be reconfigured, and (iv) which constraints must remain satisfied over time. Building on this framing, we formulate ten research challenges for the next decade, spanning theoretical guarantees for evolving systems, dynamic architectures and hybrid transitions between data-driven and model-based components, fault/anomaly-driven targeted updates, System-1/System-2 decompositions (anytime intelligence), modularity, validation under scarce labels, and evaluation protocols that quantify lifecycle efficiency and recovery/stability under drift and interventions.
\end{abstract}



\section{Edge AI is already adaptive (operationally)}
\label{sec:intro}

\EdgeAI{} is often presented as \emph{deploying smaller models under tight constraints} - compression, quantization, and hardware-aware inference.
A complementary (and increasingly dominant) view treats the edge as a \emph{runtime regime}: on-device models must operate over time under
distribution shift, heterogeneous hardware/software stacks, and time-varying budgets, and they increasingly exploit conditional computation
(routing, early exits, selective activation) as a primary efficiency mechanism \cite{wang2025empowering,scardapane2024conditional}.
In realistic deployments, the operating conditions evolve continuously:
data distributions drift (context shifts, sensor aging), resource budgets fluctuate (battery, thermals, latency), connectivity is intermittent,
and user requirements change. In short, \emph{edge systems operate under non-stationarity and time-varying constraints.}

This has immediate consequences for non-adaptive systems.
A static deployment fixes a single operating point - a model configuration, an execution policy, an implicit calibration state~- and then assumes that
the feasibility region remains stable. When conditions drift, that assumption breaks in measurable ways.
First, a fixed configuration can become \emph{infeasible}: thermal throttling, workload interference, or reduced battery headroom can push latency/energy
beyond budgets even if the model is unchanged. Second, even when execution remains feasible, predictive \emph{reliability} degrades under shift:
accuracy drops, and uncertainty may become miscalibrated, so the system may remain overconfident while wrong.
Third, long-horizon operation concentrates risk in \emph{transients} and \emph{rare time intervals} rather than in average metrics: brief regimes
(e.g., low light, motion blur, sensor drift, connectivity loss) dominate operational failures unless explicitly handled.
These failure modes motivate the central claim of this paper: sustained \EdgeAI{} cannot be reduced to static embedded inference.

\vspace{0.2em}
\noindent We therefore advance the following position statement:
\begin{quote}
\emph{\EdgeAI{} and \AdaptiveEdgeAI{} should be considered semantically equivalent as operational concepts.}
\end{quote}
If \emph{edge} implies sustained operation in the wild, then adaptivity - the ability to reconfigure computation and, when required, update model state
as conditions evolve under constraints - is not an optional add-on but a defining requirement.

This equivalence forces explicit assumptions about (i)~\emph{what changes} (data, task, hardware, user), (ii)~\emph{what is observable},
(iii)~\emph{what can be reconfigured}, and (iv)~\emph{how systems should be evaluated} over long horizons with drift and interventions.
These questions are increasingly central in the literature: surveys on test-time adaptation highlight the breadth of deployment-time shifts \cite{xiao2024beyondtta},
on-device benchmarks emphasize that adaptation must be assessed under realistic device constraints and operating conditions \cite{danilowski2025botta},
and energy/latency measurement itself requires standardized reporting to be comparable across platforms \cite{tschandl2024mlperfpower}.

\smallskip
\noindent Our goal is to align the community on a coherent agenda. We contribute:
\begin{enumerate}[leftmargin=1.2em, itemsep=0.2em]
\item \textbf{Thesis:} Static deployment is operationally incomplete for \EdgeAI{}; adaptivity is intrinsic.
\item \textbf{Framework:} A minimal \ASE{} lens that clarifies what changes, what is observed, what can be reconfigured, and what must remain satisfied over time.
\item \textbf{Agenda:} Ten research challenges for the next decade, with evaluation protocols and measurable success criteria.
\end{enumerate}


\section{From Edge AI to Adaptive Edge AI}
\label{sec:equivalence}

A useful way to interpret current practice is in terms of \emph{when} optimization happens.
Most \EdgeAI{} pipelines optimize \emph{before} deployment: models are compressed/quantized offline and then executed with fixed
architectures and operating points. Surveys of on-device AI emphasize this ``optimize-then-freeze'' workflow as the default,
with runtime adaptation typically confined to narrow mechanisms or handcrafted heuristics \cite{wang2025empowering}.

We adopt an operational definition of \EdgeAI{} as \emph{learning-enabled systems deployed on (or near) resource-constrained devices that must deliver utility over time while interacting with the physical world under time-varying budgets} \cite{wang2025empowering}.
Under this definition, a non-adaptive deployment corresponds to a fixed operating point
- a frozen configuration and execution policy - and is therefore a limiting case:
it may remain adequate over short horizons or tightly controlled regimes
(e.g., fixed industrial inspection with stable illumination and no thermal variation),
but it is operationally incomplete for sustained operation under realistic environmental and operational variability.

Three empirical properties motivate treating \EdgeAI{} and \AdaptiveEdgeAI{} as equivalent operational concepts.

\vspace{0.2em}
\noindent\textbf{(P1) Non-stationarity is the default.}
Edge data streams arise from physics and human behavior rather than curated i.i.d.\ datasets.
Even when the semantic task is stable, the effective test distribution evolves (context shifts, illumination/weather changes, motion patterns, sensor drift),
so the meaning of ``in-distribution'' changes over time.
This deployment mismatch is the explicit target of recent test-time adaptation work and on-device benchmarks \cite{xiao2024beyondtta,danilowski2025botta}.

\vspace{0.2em}
\noindent\textbf{(P2) Budgets are time-varying and multi-dimensional.}
Latency, energy, thermal headroom, bandwidth, and privacy constraints fluctuate with device state and user context.
A fixed operating point (model, precision, compute budget) cannot remain simultaneously feasible and efficient across regimes.
Thermal dynamics and throttling make sustained performance a moving target on mobile/edge hardware \cite{tan2024thermal},
and energy efficiency is now treated as a first-order benchmarking dimension across deployment scales \cite{tschandl2024mlperfpower}.

\vspace{0.2em}
\noindent\textbf{(P3) Interaction induces action-dependent streams.}
In many edge applications, the system outputs influence subsequent inputs through sensing/control decisions and user interaction.
Robots and wearables change what they observe by moving or sampling; monitoring systems trigger interventions that alter the measured process.
The resulting stream is partly endogenous to the deployed behavior, so i.i.d. assumptions can fail even in otherwise stationary environments \cite{mansfield2024activesensing_rl}.

\smallskip
\noindent Together, (P1)-(P3) imply that \EdgeAI{} requires an operational capability to
\emph{detect} meaningful change, \emph{select} feasible reconfigurations under current budgets, and \emph{apply} them safely over time.
We refer to this capability as \emph{adaptivity}; consequently, \EdgeAI{} and \AdaptiveEdgeAI{} coincide as operational concepts.
The next section formalizes the Adaptive Edge AI framework and motivates a decade-scale research agenda.

\begin{table*}[h!]
\caption{Typical instantiations of ASE variables in adaptive edge systems: observables $(o_t,c_t)\rightarrow$ actions $a_t\rightarrow$ failure modes (utility loss / violations over time).}
\label{tab:signals-knobs-risks}
\centering
\small
\begin{tabular}{p{0.31\linewidth} p{0.33\linewidth} p{0.30\linewidth}}
\toprule
\textbf{Observables $(o_t,c_t)$} & \textbf{Actions $a_t$ (reconfigure $s_t$)} & \textbf{Failure modes (over time)} \\
\midrule
Uncertainty / calibration drift & Exit depth, compute budget, model selection, abstention & Overconfidence, silent failures, missed escalations \\
Telemetry (latency/energy/thermal) & Precision scaling, caching, routing, structured sparsity & Budget overruns, instability, tail degradation \\
Shift/drift indicators (OOD, change tests) & Recalibration, memory refresh, selective updates & False alarms, oscillations, forgetting \\
Fault/anomaly indicators (sensor/hardware) & Module isolation, redundancy, fallback modes & Misattribution, cascading failures \\
Connectivity / privacy state & Local-only adaptation, offload/split execution, communication scheduling & Leakage, brittle offline behavior, stale personalization \\
\bottomrule
\end{tabular}
\end{table*}
\section{The Agent-System-Environment (ASE) lens}
\label{sec:ase}

Section~\ref{sec:equivalence} motivates \EdgeAI{} as long-horizon operation under drift and time-varying budgets, and highlights that
progress is hard to compare without time-aware protocols.
To formalize adaptivity, 
we introduce an Agent-System-Environment (ASE) lens.
ASE fixes the \emph{state variables} evolving over time - the observation and budget stream, the system configuration/state, and admissible reconfigurations~- so that
methods can be described and evaluated on common ground.

\subsection{The ASE loop (variables and roles)}
At time $t$, the environment yields observations $o_t$ and operating conditions/budgets $c_t$ (e.g., latency, energy, privacy, connectivity, thermal state).
The deployed system is in state $s_t$ (configuration, calibration, memory, and possibly parameters) and produces an output $\hat{y}_t$.
An adaptation mechanism selects an action $a_t$ that may change the system state, yielding $s_{t+1}$:
\begin{align}
&(o_t,c_t)\;\mapsto\; \hat{y}_t = f_{s_t}(o_t), \label{eq:ase_pred}\\
&a_t = \rho(o_t,s_t,c_t),\qquad s_{t+1}=g(s_t,a_t). \label{eq:ase_update}
\end{align}

\smallskip
\noindent\textbf{Environment (E).}
The data-generating process and operating regime: context shifts, sensor drift, workload/concurrency, connectivity/privacy state,
and hardware conditions (temperature, aging, faults).

\smallskip
\noindent\textbf{System (S).}
The deployed artifact implementing $f_{s_t}$: model(s), memory/caches, calibration state, and runtime configuration (precision, routing, offload),
including any fallback behavior.

\smallskip
\noindent\textbf{Adaptation mechanism (A).}
The rule $\rho$ selecting reconfigurations over time (from thresholds to learned decision rules).
Actions $a_t$ span execution-level reconfiguration (conditional routing/early exits, precision scaling) and state-level updates
(cache/memory refresh, recalibration, selective adaptor/parameter updates, offload/splitting) \cite{scardapane2024conditional,jazbec2024fastsafe}.

\smallskip
\noindent The ASE loop can be viewed as a constrained POMDP specialized to the edge setting:
state, action, and observation spaces are instantiated around device-level reconfigurations
and time-varying hardware and connectivity budgets, fixing a concrete deployment vocabulary
without requiring the full generality and computational requirements of unconstrained POMDPs.

\smallskip
\noindent Learning under non-stationarity with tight time and memory constraints and delayed or missing labels is well-studied in data stream learning~\cite{gama2014survey, losing2018incremental}.
ASE subsumes this as a special case (fixed $s_t$, actions limited to parameter updates) and extends it with a multi-dimensional constraint stream~$c_t$ and a rich action space~$a_t$ that includes execution-level reconfigurations (routing, precision, offload) alongside model updates.

\subsection{Net utility under budgets (execution + reconfiguration)}
ASE makes explicit that adaptivity should be evaluated as a \emph{net} gain under budgets: it consumes the same resource budget as inference.
Over a horizon $T$, a generic constrained objective reads
$
\max_{\rho}\;\; \mathbb{E}\Big[\sum_{t=1}^{T} u(\hat{y}_t,y_t)-\lambda\cdot \mathcal{C}_{\text{total}}(s_t,a_t,c_t)\Big],
\label{eq:objective}
$
subject to time-varying constraint violations
$
\mathbb{E}\Big[\sum_{t=1}^{T} \mathrm{viol}(c_t)\Big]\le \epsilon_t,
\label{eq:constraint}
$
where labels $y_t$ may be delayed or missing.

\smallskip
\noindent We separate the cost of running the current configuration from the overhead of changing it:
\begin{equation}
\mathcal{C}_{\text{total}}=
\underbrace{\mathcal{C}_{\text{run}}(s_t,c_t)}_{\text{execution}}
+\underbrace{\mathcal{C}_{\text{chg}}(s_t,a_t,c_t)}_{\text{reconfiguration overhead}}.
\label{eq:cost}
\end{equation}
This separation is operationally relevant in on-device settings where adaptation is periodic, resource-limited, and measured end-to-end \cite{danilowski2025botta},
and it aligns with benchmarking efforts that emphasize standardized energy accounting to ensure cross-platform comparability \cite{tschandl2024mlperfpower}.

\smallskip
\noindent\textbf{Constraint violations (illustrative).}
One simple instantiation is a sum of budget exceedances:
$
\mathrm{viol}(c_t)\,=\,
\mathbbm{1}[\mathrm{lat}_t > b^{\mathrm{lat}}_t]
+\mathbbm{1}[\mathrm{en}_t > b^{\mathrm{en}}_t]\nonumber 
+\mathbbm{1}[\mathrm{bw}_t > b^{\mathrm{bw}}_t]
+\mathbbm{1}[\mathrm{priv}_t < b^{\mathrm{priv}}_t],
$
where budgets~$b_t$ may vary over time with device state and context.

\subsection{Instantiating $(o_t,c_t)$ and $a_t$ in practice}
\label{sec:ase_instantiation}

In deployed edge systems, the stream $(o_t,c_t)$ in~\eqref{eq:ase_pred}-\eqref{eq:ase_update} is realized through
data-side observables (features, uncertainty, calibration error), and system-side telemetry (latency/energy/thermal state, connectivity/privacy regime),
while $a_t$ corresponds to a restricted set of admissible reconfigurations that modify $s_t$ under budgets.
Table~\ref{tab:signals-knobs-risks} summarizes common instantiations of these quantities and the corresponding failure modes that must be controlled
over time. With $(o_t,c_t,s_t,a_t)$ explicit, the open problems in Section~\ref{sec:challenges} can now be stated as requirements on observability, admissible actions, and long-horizon control of violations~\cite{wang2025empowering,danilowski2025botta,scardapane2024conditional}.

\section{Ten challenges for the next decade}
\label{sec:challenges}

We outline ten algorithmic and systems challenges that follow from the ASE view of~\EdgeAI{} as operation over time under drift and constraints.
Each challenge is stated as: Problem $\rightarrow$ Core difficulty $\rightarrow$ Evaluation protocol $\rightarrow$ Success.

\subsection*{C1: Can we provide theoretical guarantees for adaptive edge systems under drift?}
\noindent
\textbf{Research question.}
Can we provide asymptotic and non-asymptotic theoretical guarantees on \emph{risk} and \emph{constraint violations} when the deployed configuration $s_t$
(and possibly parameters) evolves online under non-stationarity?

\noindent
\textbf{Problem.}
Develop guarantees for edge operation over time that jointly cover:
(i)~risk control under shift (e.g., PAC/PAC-Bayes-style bounds and domain-adaptation bounds) \cite{seeger2002,hennequin2024pacbayes_da_mv,wang2024fdivergence_da} 
(ii)~non-stationary online-learning guarantees with explicit drift dependence (dynamic/adaptive/interval regret) \cite{zinkevich2003,hazan2009,zhao2025nonstationary_ol},
and (iii)~constraint-aware guarantees that bound budget violations over time (latency/energy/safety/privacy) \cite{hutchinson2024safe_oco,lechowicz2024chasing_ltc}.

\noindent
\textbf{Core difficulty.}
In \EdgeAI{}, the learning problem is not only non-i.i.d.; it is often \emph{policy-coupled}:
reconfiguration actions~$a_t$ (routing, sampling, offload, abstention) affect what is observed and which feedback becomes available,
yielding action-dependent data and delayed/partial supervision.
As a consequence, both the loss process and the information structure evolve with $(s_t,a_t)$, not only with exogenous drift \cite{suk2024smooth_nonstationary_bandits}.

\noindent
\textbf{Evaluation protocol.}
Report risk and constraint violations as \emph{time series} over long horizons, explicitly marking drift and update windows.
In addition to global averages, include worst-interval / worst-slice summaries aligned with non-stationary regret notions \cite{zhao2025nonstationary_ol}.
When uncertainty sets are used to trigger or accept updates, report their online validity under shift (e.g., coverage/control via online conformal prediction) \cite{gibbs2024onlineconformal}.

\noindent
\textbf{Success.}
Guarantees with explicit dependence on drift/variation and constraint models (e.g., variation budgets, path length, long-term constraints),
together with empirical evidence of controlled risk and bounded violation rates under declared drift/intervention scripts.

\subsection*{C2: Can an edge system switch paradigms without  breaking?}
\noindent
\textbf{Research question.}
Can we switch (or hybridize) \emph{learned} and \emph{structured/physics- or rule-based} components online - i.e., change the model class inside $s_t$ via actions $a_t$ - while keeping utility and budget compliance predictable over time?

\noindent
\textbf{Problem.}
Enable reconfiguration actions $a_t$ that modify the system state $s_t$ not only in \emph{execution} (routing, depth, precision), but also in the
\emph{computational model} itself: selecting among learned modules, structured estimators/controllers, physics- or rule-based components, or hybrid compositions, depending on
$(o_t,c_t)$ (e.g., label rate, physics mismatch, connectivity, safety regime) \cite{scardapane2024conditional,quarteroni2025sciml,danilowski2025botta}.

\noindent
\textbf{Core difficulty.}
Paradigm switches are discrete. A transition changes both the realized mapping $f_{s_t}$ and the internal state carried across time
(calibration, latent/memory state, estimator/controller state). Without an explicit \emph{state-transfer} specification
(e.g., $s_{t+1} \leftarrow \mathrm{T}_{a_t}(s_t,o_t,c_t)$) and a switching cost model $\mathcal{C}_{\mathrm{chg}}(s_t,a_t,c_t)$,
systems may exhibit transient violations (latency/energy spikes, unsafe outputs) or \emph{chattering} (oscillatory switching) under noisy observables and tight budgets.

\noindent
\textbf{Evaluation protocol.}
Use benchmark streams with declared \emph{mode-change events} in $(o_t,c_t)$: budget shifts, connectivity drops, changes in label availability,
and physics-mismatch regimes. Report:
(i) utility-cost trajectories before/during/after each switch;
(ii) time-to-recover (return-to-spec time) and failure probability per switch;
(iii) constraint violations in the transition window;
(iv) switching frequency and hysteresis (to quantify chattering).
When switching is gated by uncertainty or exit criteria, include risk-controlled gating variants \cite{jazbec2024fastsafe}.
For edge settings with split/offload options, include benchmarks where switching couples with placement decisions \cite{colocrese2024mdiexit}.

\noindent
\textbf{Success.}
Mode changes with bounded transient degradation and bounded violation probability, stable long-horizon behavior (no chattering),
and reproducible trade-offs across operating conditions and devices.

\subsection*{C3: When utility drops, can we localize the cause and fix only what is broken?}
\noindent
\textbf{Research question.}
Given the stream $(o_t,c_t)$ and the current system state $s_t$, can an edge system infer \emph{why} utility or constraint compliance degraded
(sensor fault, regime shift, component mismatch, or budget shift), and then choose the \emph{smallest effective} action $a_t$ that restores
utility while keeping violations bounded?

\noindent
\textbf{Problem.}
Move from anomaly \emph{detection} to anomaly \emph{attribution} and \emph{targeted repair}.
Anomaly detection on streams is now well-studied and benchmarked~\cite{liu2024tsbad,darban2024dltsad},
but adaptive \EdgeAI{} additionally needs: (i)~identifying the fault source (which part of $E$ or $S$ changed),
and (ii)~selecting a localized intervention (recalibration, memory refresh, module swap, limited parameter/adaptor update) rather than global retraining.

\noindent
\textbf{Core difficulty.}
Attribution is ill-posed under partial observability: multiple causes can co-occur, anomalies propagate across variables and modules,
and the same drop in $u(\hat y_t,y_t)$ (or rise in $\mathrm{viol}(c_t)$) can be explained by different latent changes.
Moreover, interventions themselves can confound diagnosis by altering the subsequent stream.
Recent work on temporal/causal root-cause analysis makes progress on structured attribution \cite{rehak2024temporalrca,han2025aerca},
but integrating attribution with \emph{minimal} interventions that are edge-feasible (global retrain is often infeasible on-device) remains largely open. 

\noindent
\textbf{Evaluation protocol.}
Use long-horizon streams with \emph{labeled fault sources} and controlled injections (sensor bias/drift/dropout, packet loss,
constraint-regime switches, component failures). Report:
(i) detection delay; (ii) attribution precision/recall over fault sources;
(iii) \emph{intervention minimality} (changed parameters/modules; energy/time) and \emph{net} regained utility under budgets;
(iv) collateral effects on unaffected slices.
Where repair is claimed, include minimal-change baselines and (when possible) verifiable post-repair properties \cite{fu2024reglo}.

\noindent
\textbf{Success.}
Near-real-time recovery through \emph{localized} actions whose effect is attributable and stable over time:
bounded violations, predictable side-effects, and measurable net gain (utility regained minus intervention cost),
without frequent full retraining or global re-optimization, aligning with 
minimal-change principles such as the ones studied in neural network repair/model update methods \cite{chen2024inner}.

\subsection*{C4: Can we make "spend more compute only when necessary" provably safe at the edge?}
\noindent
\textbf{Research question.}
Can we design a two-tier (or multi-tier) system where a cheap pathway handles most inputs, and an expensive pathway is invoked only when needed,
so that the resulting policy~$a_t$ (escalate or not) yields predictable \emph{tail-risk} reduction under drift while respecting time-varying budgets~$c_t$?

\noindent
\textbf{Problem.}
Construct \emph{anytime} systems by decomposing the deployed state $s_t$ into fast and slow components (``System~1'' and ``System~2''),
and enabling actions $a_t$ that allocate computation on demand (e.g., early exits, conditional routing, selective activation) \cite{teerapittayanon2016,scardapane2024conditional}.
The goal is not lower average FLOPs per se, but reliability under constraints: spend additional compute only on inputs and regimes where it measurably reduces risk.

\noindent
\textbf{Core difficulty.}
Escalation is simultaneously a \emph{reliability} decision and a \emph{budget} decision.
Under shift, naive difficulty/uncertainty scores can become miscalibrated, causing missed escalations precisely on hard/OOD time intervals;
under tight budgets, overly conservative gating collapses into frequent escalations and erases savings.
Technically, the challenge is to produce uncertainty signals that remain meaningful.
Recent work connects early exiting to distribution-free risk control and to \emph{nested} (sequentially consistent) prediction sets, enabling controlled escalation policies~\cite{jazbec2024fastsafe,jazbec2024nested}.

\noindent
\textbf{Evaluation protocol.}
Report performance as \emph{risk-cost trade-offs} over time.
At minimum include:
(i) tail-risk at fixed \emph{average} latency/energy;
(ii) escalation frequency and escalation errors (missed vs unnecessary escalations);
(iii) constraint violations during drift regimes (time series, not only steady state).
When risk control is claimed, report risk-coverage style curves and risk-controlled operating points rather than only accuracy-FLOPs plots \cite{jazbec2024fastsafe,goren2024hsc}.

\noindent
\textbf{Success.}
Stable long-horizon reduction in tail risk with bounded violation rates, achieved through escalation policies that remain calibrated under drift and do not
exhibit chattering or regime-dependent pathologies.

\subsection*{C5: Can we compose and update modular edge intelligence without breaking the whole system?}
\noindent
\textbf{Research question.}
Can we represent the deployed state $s_t$ as a \emph{set of composable modules} - including \emph{memory} - that can be added or updated locally
(via actions $a_t$) while preserving predictable system-level behavior and budget compliance?

\noindent
\textbf{Problem.}
Move from monolithic models toward modular edge systems: libraries of experts/adapters/skills plus explicit memory components (episodic buffers, retrieval
indices, caches) that support long-horizon operation and context specialization.
In this view,~$s_t$ contains multiple interacting subsystems (compute modules and memory state), and adaptation actions~$a_t$ include
adding/removing modules, updating a single module, or refreshing/compacting memory, rather than retraining the full stack
\cite{hu2024smn,liu2024s6mod,liu2025psec}.
This parallels the functional motivation for specialization in biological systems (distinct subsystems for rapid recall vs slower consolidation).

\noindent
\textbf{Core difficulty.}
Changing one module can shift hidden representations, calibration, routing decisions, or retrieval distributions.
Memory exacerbates this: retrieval-based behavior is distribution-dependent, and local changes to the embedding/key space can invalidate stored items.
Therefore, modular updates require explicit \emph{compatibility mechanisms} 
and lightweight verification to prevent local fixes that degrade global utility or violate constraints.

\noindent
\textbf{Evaluation protocol.}
Use \emph{module-isolation} tests aligned with ASE:
apply actions that modify exactly one component of~$s_t$ (swap/update a single module, add a new skill, or refresh/compact memory) while freezing the rest.
Report:
(i)~system-level utility and constraint violations over time;
(ii)~changes in performance metrics on unaffected tasks/slices (including calibration and abstention behavior);
(iii)~module-level accountability (attribution of gains/regressions to the modified component);
(iv)~memory-specific metrics when present (retrieval hit-rate / staleness, memory growth, and energy/time per memory maintenance step).

\noindent
\textbf{Success.}
Local updates with predictable global effects: improvements persist, deteriorations are bounded and detectable, and modules (including memory representations) remain reusable across contexts. 

\subsection*{C6: Can we trigger, accept, or rollback adaptation without ground truth?}
\noindent
\textbf{Research question.}
When labels are delayed, sparse, or absent, can we decide \emph{when} to adapt and \emph{whether} an update should be kept (or rolled back),
while controlling risk and constraint violations over time?

\noindent
\textbf{Problem.}
In the ASE loop, supervision is part of the environment: $y_t$ may arrive late, intermittently, or under a fixed labeling budget.
The system must therefore solve two coupled decisions:
(i)~\emph{triggering} (choose $a_t$ that initiates an adaptation step), and
(ii)~\emph{acceptance} (keep/rollback a modified state~$s_{t+1}$).
This includes edge-feasible mechanisms that \emph{change the feedback process} 
such as selective querying (active learning) or deferral,
and mechanisms that \emph{substitute} supervision with proxy objectives (self-/semi-supervised signals) when labels are unavailable. 

\noindent
\textbf{Core difficulty.}
Without ground truth, acceptance relies on proxy signals (self-supervised losses, entropy, reconstruction error, agreement across exits/models)
that can become miscalibrated under drift.
Moreover, unsupervised test-time adaptation can create feedback loops (confirmation bias) that silently degrade reliability, so ``improvement''
cannot be assumed from decreasing proxy losses alone \cite{xiao2024beyondtta}.
Active querying helps but is itself constrained (cost, privacy, latency, user burden) and must be allocated where it most reduces risk.

\noindent
\textbf{Evaluation protocol.}
Use streams with explicit label-delay processes and/or fixed label budgets; include conditions where only a small calibration window is available.
Report:
(i)~risk calibration under drift;
(ii)~false-trigger rate (adapt when unnecessary) and miss rate (fail to adapt when needed);
(iii)~acceptance errors (keep a harmful update vs rollback a beneficial one) and the cost of rollback;
(iv)~utility and constraint violations as \emph{time series} across drift/update windows.
Include abstention/selective-prediction baselines as safety mechanisms under shift \cite{liang2024scshift,pugnana2024scbench}.

\noindent
\textbf{Success.}
Update-trigger and acceptance criteria with finite-sample, distribution-free (finite-sample) control of risk/violations from small calibration windows,
remaining valid in online/non-stationary settings - e.g., via conformal risk control and online conformal inference \cite{angelopoulos2024crc,gibbs2024onlineconformal}.

\subsection*{C7: Is \emph{efficient inference} irrelevant if maintenance dominates?}
\noindent
\textbf{Research question.}
Over a long horizon, can we minimize the \emph{total} cost of operating an adaptive edge system
\(\sum_t \mathcal{C}_{\text{run}}(s_t,c_t) + \mathcal{C}_{\text{chg}}(s_t,a_t,c_t)\)
while keeping utility above specification and constraint violations bounded?

\noindent
\textbf{Problem.}
Shift the objective of \emph{efficiency} from static inference FLOPs to \emph{lifecycle-efficient operation}:
training, deployment, monitoring, periodic adaptation/maintenance, and (when relevant) communication.
In ASE terms, this is precisely the cost term in~\eqref{eq:cost}: the system must be efficient at executing~$f_{s_t}$ and also at deciding and applying changes to~$s_t$ over time.

\noindent
\textbf{Core difficulty.}
In realistic deployments, the dominant cost can move from execution to \emph{maintenance}:
monitoring, periodic adaptation/personalization, and distributed update mechanisms can exceed inference energy/time over the horizon.
Moreover, energy/latency accounting is platform-dependent and easy to misreport without standardized methodology,
making comparisons across heterogeneous hardware and operating states unreliable \cite{tschandl2024mlperfpower,barbierato2024greenai}.
This is amplified on-device, where background load, thermal throttling, and OS scheduling affect both $\mathcal{C}_{\text{run}}$ and $\mathcal{C}_{\text{chg}}$
\cite{wang2025empowering}.

\noindent
\textbf{Evaluation protocol.}
Report \emph{horizon-level} metrics:
total energy/time to keep utility above a target (or within a risk bound) \emph{and} bounded tail risk, including monitoring and update costs.
Disclose measurement methodology (power logging source, integration window, device state, thermal regime) for reproducibility \cite{tschandl2024mlperfpower}.
In addition to totals, report an amortized metric such as Joules per unit of maintained/recovered utility over the horizon, and separate
\(\sum_t \mathcal{C}_{\text{run}}\) from \(\sum_t \mathcal{C}_{\text{chg}}\).

\noindent
\textbf{Success.}
Methods and systems that achieve lower lifecycle energy/time with stable long-horizon behavior, demonstrated on resource-constrained devices,
with explicit separation of execution vs maintenance costs and clear evidence that the net gain persists under realistic operating conditions
\cite{li2024etuner,wang2025empowering}.

\subsection*{C8: Can we adapt privately when the network is unreliable?}
\noindent
\textbf{Research question.}
When privacy and connectivity are part of the constraint stream $c_t$, can an edge system personalize and adapt 
using only admissible observables and actions, while remaining stable through extended offline periods?

\noindent
\textbf{Problem.}
Enable personalization and continual adaptation when raw data cannot leave the device and coordination is intermittent.
In ASE terms, privacy and connectivity constrain both the information available for learning (what feedback can be used or shared)
and the action space (which updates are permissible and when synchronization can occur), favoring local-only updates
(e.g., adapters) with intermittent aggregation when communication is available~\cite{wang2024fclsurvey,li2024localadapt}.

\noindent
\textbf{Core difficulty.}
Privacy limits what statistics/representations/gradients can be communicated; connectivity induces asynchronous, bursty synchronization.
The system must therefore (i) avoid unstable personalization during long offline intervals (overfitting/forgetting),
(ii) prevent leakage through updates or stored state, and (iii) remain robust to staleness and heterogeneity when aggregation resumes
\cite{li2024localadapt,wang2024fclsurvey}.
These constraints are coupled: more aggressive local adaptation increases drift from shared baselines and complicates safe re-alignment.

\noindent
\textbf{Evaluation protocol.}
Use realistic connectivity traces (online/offline schedules) and explicit communication budgets.
Report: (i) utility vs communication (bytes/rounds) and local compute/energy; (ii) decay and recovery across offline intervals (time series);
(iii) stability under heterogeneity when synchronization resumes; (iv) privacy reporting consistent with the threat model
(e.g., DP accounting if claimed, or explicit attack evaluation otherwise). Include baselines separating local-only adaptation from intermittent
coordination \cite{wang2024fclsurvey}.

\noindent
\textbf{Success.}
Competitive personalization with minimal communication and stable long-horizon behavior under realistic offline durations,
with privacy guarantees (or empirically validated protections) under an explicit threat model.

\subsection*{C9: Can we keep guarantees when the hardware becomes part of the drift?}
\noindent
\textbf{Research question.}
When the compute substrate and sensing stack evolve over time, can we maintain predictable utility and bounded violations by adapting $s_t$
(and selecting mitigation actions $a_t$) under tight budgets?

\noindent
\textbf{Problem.}
Maintain utility and constraint compliance when the \emph{compute substrate itself} changes with conditions and aging, including thermal effects
(temperature dependence, DVFS/throttling), voltage-margin trimming/undervolting, soft errors in memory/compute, wear-out and drift, and non-idealities
in emerging accelerators (e.g., analog noise and conductance drift in in-memory compute)
\cite{tan2024thermal,ahmadilivani2024hardware,bolchini2024resilience,rasch2024aimc_training}.
In ASE terms, this is a coupled drift: hardware affects the realized mapping $f_{s_t}$, the constraint stream $c_t$ (latency/energy regimes),
and sometimes the observation stream $o_t$ (sensor/ADC drift).

\noindent
\textbf{Core difficulty.}
Hardware variability induces \emph{state-dependent}: the same $s_t$ and $o_t$ can produce different outputs depending on
temperature/voltage/aging state, breaking the assumption that the deployed model is a fixed function plus i.i.d.\ noise.
Failures can be silent (Silent Data Corruption, SDC), interact with conditional execution (different routes stress different units), and occur precisely when budgets are tight, while edge constraints limit frequent recalibration; robust operation therefore couples learning, calibration, and device/accelerator characterization
\cite{yu2024survey_fault_injection_ai,chen2024wages}.

\noindent
\textbf{Evaluation protocol.}
Run long-horizon stress tests that sweep temperature/voltage/frequency, induce throttling, and inject controlled faults/drift in memory and compute.
Report: (i)~utility trajectories and tail-risk, including SDC rate where applicable; (ii)~constraint violations under throttling and degraded regimes;
(iii)~robustness under specified fault/drift injection models; (iv)~frequency and \emph{cost} (energy/time) of mitigation actions (error detection, redundancy,
recalibration/compensation) \cite{yu2024survey_fault_injection_ai,zheng2025save}. When analog/IMC is involved, include drift profiles and
compensation overhead \cite{rasch2024aimc_training}.

\noindent
\textbf{Success.}
Sustained utility with bounded violations under realistic variability and degradation, with low overhead and predictable degradation modes:
lightweight compensation that tracks substrate drift (and its induced~$c_t$ changes) without frequent full retraining
\cite{katti2024bayes2imc,martemucci2025ferroelectric_memristor}.

\subsection*{C10: What does it mean to \emph{win} when systems adapt over time?}
\noindent
\textbf{Research question.}
Which protocols and metrics make adaptive edge systems comparable when $(o_t,c_t)$, $s_t$, and permissible actions $a_t$ evolve over time?

\noindent
\textbf{Problem.}
Static test sets and single-shot reporting cannot characterize adaptive behavior:
the relevant phenomena are \emph{transient} and \emph{path-dependent} - recovery after drift, oscillations under repeated interventions,
and time-varying budget violations.
In ASE terms, \emph{performance} is not a scalar but a trajectory of utility and constraint satisfaction under a declared stream and adaptation rules.

\noindent
\textbf{Core difficulty.}
Reported gains depend strongly on choices that are often under-specified:
(i) the \emph{shift script} (type/severity/timing of changes in $(o_t,c_t)$),
(ii) the \emph{adaptation schedule} (continuous vs periodic; what data/labels are available and when),
and (iii) the \emph{cost accounting} (what is included in energy/latency and how it is measured on-device).
Recent on-device adaptation benchmarks emphasize that these details can dominate conclusions and must be reported explicitly \cite{danilowski2025botta,fan2024benchmarking},
while system-level benchmarking efforts stress standardized measurement methodology to make cross-platform comparisons meaningful \cite{tschandl2024mlperfpower}.

\noindent
\textbf{Evaluation protocol.}
Evaluate \emph{trajectories over time}, not only the post-adaptation steady state.
Report:
(i) utility~$U(t)$ (including calibration measures when relevant),
(ii)~cost~$\mathcal{C}_{\text{run}}(t)$ and~$\mathcal{C}_{\text{chg}}(t)$ (execution vs reconfiguration),
and (iii) violation indicators~$V(t)$ derived from $\mathrm{viol}(c_t)$.

Include recovery and stability metrics, e.g.:
(i) \textit{Energy-to-Recover (E2R)}: Joules from shift onset (or detection) to regain within~$\delta$ of pre-shift utility;
(ii) \textit{Time-to-Recover (T2R)}: time/frames to recover to the same threshold;
(iii) \textit{Stability Score (SS)}: oscillation/variance of~$U(t)$ and~$V(t)$ under repeated shifts and interventions.
When reporting energy/latency, disclose the measurement protocol (instrumentation, integration window, phases included, device state/thermal regime),
following reproducible benchmarking best practices \cite{tschandl2024mlperfpower,bartoli2025benchmarking}.
\noindent
\textbf{Success.}
Community benchmarks and reporting standards that fix: (i) declared shift/intervention scripts, (ii) recovery/stability metrics, and
(iii) on-device cost accounting - so that adaptive edge systems are comparable across algorithms, hardware, and operating regimes
\cite{danilowski2025botta,fan2024benchmarking,tschandl2024mlperfpower}.

\begin{table}[t]
\caption{Minimal checklist for adaptive edge AI experiments (to enable comparability).}
\label{tab:checklist}
\centering
\small
\begin{tabular}{p{0.94\linewidth}}
\toprule
\textbf{Report (at minimum):}\\
\midrule
\textbf{(D) Drift/interventions:} what changes (data/task/hardware/user), severity/timescale, and the declared shift/intervention
(change points known/unknown, staged/gradual, injected faults).\\
\textbf{(C) Constraints:} latency/energy/memory/privacy/connectivity budgets and how $c_t$ varies over time (device state, throttling, offline periods).\\
\textbf{(A) Actions:} admissible $a_t$ (routing, precision, memory, params, offload), adaptation schedule (continuous/periodic), rollback/fallback policy.\\
\textbf{(S) Signals:} observables used to trigger/select actions (uncertainty, drift tests, telemetry) and their thresholds/costs.\\
\textbf{(M) Metrics as time series:} $U(t)$, $\mathcal{C}_{\text{run}}(t)$, $\mathcal{C}_{\text{chg}}(t)$, and $V(t)$; include recovery/stability summaries when relevant.\\
\textbf{(P) Protocol:} time-aware splits; online vs offline; label availability and delay/budget model; data used for adaptation vs evaluation.\\
\textbf{(B) Baselines:} frozen deployment; oracle/upper bounds where meaningful; ablations isolating each action type.\\
\bottomrule
\end{tabular}
\end{table}

\paragraph*{Reporting standard for adaptive Edge AI}
\label{sec:evaluation}

Challenge~C10 calls for comparability under drift and intervention scripts and explicit on-device cost reporting.
As a concrete step, we propose the \emph{reporting standard} in Table~\ref{tab:checklist}: it specifies the drift and constraints regime,
the admissible adaptation actions, and the time-series metrics required to interpret results beyond static accuracy.

\section{Conclusion}
\label{sec:conclusion}

\EdgeAI{} requires long-horizon operation in the wild: the data stream, compute budgets, connectivity, and even the compute substrate vary over time.
Under these conditions, a fixed deployment checkpoint inevitably drifts toward either \emph{utility failure} (miscalibration and risk growth under shift)
or \emph{budget failure} (latency or energy violations under changing device state), often with silent degradation.
This motivates our operational thesis that \EdgeAI{} and \AdaptiveEdgeAI{} coincide in practice.

The ASE lens proposed in this paper makes this thesis precise by exposing the minimal objects that determine adaptive behavior over time:
the observable stream~$(o_t,c_t)$, the evolving system state~$s_t$, and the admissible interventions~$a_t$ under constraints.
Within this coordinate system, we formulate ten research questions for the next decade - from guarantees under drift to modularity, private/offline adaptation,
hardware variability, and time-series evaluation protocols - and we propose a minimal reporting standard to make progress comparable across algorithms,
devices, and operating regimes.

\section*{Acknowledgments}
This paper is supported by PNRR-PE-AI FAIR project
funded by the NextGeneration EU program.

\bibliography{references}
\bibliographystyle{IEEEtran}

\end{document}